# Comment on "Exact-corrected confidence interval for risk difference in noninferiority binomial trials"

*Antonio Martín Andrés[1*] and Inmaculada Herranz Tejedor[2]*

[1] Bioestadística, Facultad de Medicina, Universidad de Granada, Granada, Spain..
[2] Bioestadística, Facultad de Medicina, Universidad Complutense de Madrid, Madrid, Spain.

## ABSTRACT

The article by Hawila & Berg (2023) that is going to be commented presents four relevant problems, apart from other less important ones that are also cited. First, the title is incorrect, since it leads readers to believe that the confidence interval defined is exact when in fact it is asymptotic. Second, contrary to what is assumed by the authors of the article, the statistic that they define is not monotonic in delta. But it is fundamental that this property is verified, as the authors themselves recognize. Third, the inferences provided by the confidence interval proposed may be incoherent, which could lead the scientific community to reach incorrect conclusions in any practical application. For example, for fixed data it might happen that a certain delta value is within the 90%-confidence interval, but outside the 95%-confidence interval. Fourth, the authors do not validate its statistic through a simulation with diverse (and credible) values of the parameters involved. In fact, one of its two examples is for an alpha error of 70%!

## 1.- Introduction.

Hawila & Berg (2023) have made a considerable effort to propose a novel confidence interval (CI) for the risk difference in non-inferiority binomial trials which they call "EC $\delta$-projected Z-score". The proposed CI is dependent on the prespecified noninferiority margin $\delta_0$ and is consistent with the exact unconditional test of Chan (1998) for $\delta_0 \geq 0$. Nevertheless, we believe that it is necessary to complete and/or further explain different aspects of their article related to some statements or definitions that may cause mistakes, some missing bibliographical reference, and the incoherence and non-monotony of the inference method proposed. From now on, we will use the same notation as the authors, although on some occasions an extra notation will be added.

---
[*] Correspondence to: Bioestadística, Facultad de Medicina, C8-01, Universidad de Granada, 18071 Granada, Spain. Email: amartina@ugr.es. Phone: 34-58-244080.

**2. Some problems detected.**

The first problem is in the title of the paper ("Exact-correct confidence interval…"), as it may lead readers to think that an exact CI is being proposed. It is true that the CI proposed guarantees the coverage of $-\delta_0$, since that value is obtained through the inversion of the exact test of Chan (1998). However, the coverage of all of the rest of $\delta$ values of the CI is not guaranteed, since they are obtained through the inversion of the asymptotic test proposed -based on the statistic $Z_\delta^{EC}(X_T, X_C)$- and they are not obtained through the inversion of an exact test.

Another aspect that may confuse readers is related to the two different definitions of $\delta$. In the Introduction to Section 2, in Table 1 and in Section 2.1, $\delta = P_T - P_C$ is defined, but from Section 2.2 onwards it is assumed that $-\delta = P_T - P_C$.

Regarding the statistic $Z_\delta(X_T, X_C)$, which is the basis of the method which the authors call "$\delta$-projected $Z$-score", it is necessary to further explain several questions. Firstly, the statistic was proposed by Mee (1984), although it was Miettinen & Nurminen (1985) who provided the explicit formula of the estimator of the nuisance parameter $P_C$; in fact, the statistic of M&N is $[(n_T+n_C)/(n_T+n_C-1)]^{0.5} \times Z_\delta(X_T, X_C)$. Secondly, the theoretical demonstrations that the statistic verifies the two properties of convexity of Barnard and the property of monotonicity in $\delta$ were made by Martín Andrés & Herranz Tejedor (2004).

Röhmel (2005) gave a more detailed demonstration of the former, whereas Herranz Tejedor & Martín Andrés (2008) demonstrated that if one of the two properties of convexity is verified, then the other one must necessarily be verified.

The exact test of Chan (1998) -which is based on the order provided by the statistic $Z_\delta(X_T, X_C)$ of expression (9) of H&B-, has the problem that its *p*-values $p^{exact}(x_T, x_C)$ may not be monotonic in $\delta$ (Röhmel & Mansmann 1999), as can be seen in Figure 1 of H&B. This means that the inversion of the test may provide a confidence bound with some gaps, which is why Chan (1999) proposed obtaining the CI filling in these gaps; eventually, the test for $\delta_0$ can be made based on the CI obtained in this way. This is the procedure that gives rise to the *p*-value $p^{ZC}(x_T, x_C)$ of the paper. However, H&B proposed a CI($\delta_0$) which is based on the value of $p^{exact}(x_T, x_C)$ in $\delta_0$ - $p^{exact}_{\delta_0}(x_T, x_C)$ -, so that $-\delta_0 \notin \mathrm{CI}(\delta_0) \Leftrightarrow p^{exact}_{\delta_0}(x_T, x_C) < \alpha/2$. This procedure means a return to the problem highlighted by Röhmel & Mansmann (1999): the inferences obtained may be incoherent. Indeed, if $-\overline{\delta}_0 < -\delta_0$ and $p^{exact}_{\delta_0}(x_T, x_C) < \alpha/2 < p^{exact}_{\overline{\delta}_0}(x_T, x_C)$ -as occurs with some values of Figure 1 of H&B- then we would reject the null hypothesis for the non-inferiority margin $\delta_0$, but we would not reject it for the non-inferiority margin $\overline{\delta}_0 > \delta_0$.

The statistic of expression (20) used by H&B can be written in the following alternative format:

$$Z^{EC}_\delta(X_T, X_C) = \frac{\hat{d}_0 + \delta}{\hat{\sigma}_\delta} \text{ where } \hat{d}_0 = -\delta_0 + \hat{\sigma}_{\delta_0} \Phi^{-1}\{1 - p^{exact}(X_T, X_C)\},$$

In that expression $\hat{d}_0$, which is the only new term, refers to the value that $\hat{d} = \hat{P}_T - \hat{P}_C$ should have so that $p^{exact}(X_T, X_C) = p^{asy}(X_T, X_C)$; this term is constant for some set values of ($n_T$, $n_C$, $X_T$, $X_C$, $\delta_0$). From the previous expression it is easy to see that $Z^{EC}_\delta(X_T, X_C) < Z_\delta(X_T, X_C) \Leftrightarrow$ $\hat{d}_0 < \hat{d}$, which also happens if and only if the asymptotic test of Mee is liberal in relation to the exact test of Chan i.e. $\Leftrightarrow p^{asy}(X_T, X_C) < p^{exact}(X_T, X_C)$. The latter is what occurs in almost all of the settings ($n_T$, $n_C$, $X_T$, $X_C$, $\delta_0$). Nevertheless, the statistic $Z_\delta^{EC}(X_T, X_C)$ is not always monotonic in $\delta$, although it almost always is. For example, for ($n_T$=6, $n_C$=6, $x_T$=6, $x_C$=0, $\delta_0$=0.05), it is obtained that $p^{exact}$=0.00013, $\hat{d}_0$=1.0019 > $\hat{d}$=1 (the asymptotic test of Mee is conservative) and $Z_\delta^{EC}(x_T, x_C)$ is not monotonic in $\delta$, since $Z_{\delta=-0.9900}^{EC}$=0.2921, $Z_{\delta=-0.9981}^{EC}$=0.2133 and $Z_{\delta=-0.9990}^{EC}$=0.2242. This occurs on more occasions, since it can be demonstrated that for the previous setting $Z_\delta^{EC}(X_T, X_C)$ reaches a minimum in $\delta=-1/\hat{d}_0=-0.9981$.

The authors point out in their Discussion that "*It is natural to expect confidence intervals for the effect size $\delta$ not to depend on noninferiority design $\delta_0$. However, the Chan exact p-value requires the specification of $\delta_0$ ... Therefore, any confidence interval method that is consistent with the Chan exact p-value will naturally be dependent on $\delta_0$*". In our opinion, the first statement is totally correct. This is what happens with all of the CIs obtained through the inversion of a test; e.g. the CI for the average $\mu$ of a normal distribution, does not depend on the value of $\mu_0$ which one might have in mind for the test whose null hypothesis is H: $\mu=\mu_0$. The second statement should be more fully explained. To carry out the exact test of Chan, it is necessary to specify a maximum value $-\bar{\delta}_0$ of $P_T-P_C$ under the null hypothesis. Therefore, if we call it $-\delta_0$ that does not mean that $\delta_0$ is the value that one has in mind as the prespecified noninferiority margin. That is why the third statement is incorrect: the inversion of the exact test of Chan does not depend on noninferiority design $\delta_0$.

Finally, the authors carry out some comparisons between the CIs provided by the asymptotic methods named by EC (new), MN (Miettinen & Nurminen) and Wald, as well as the exact method of CZ (Chan & Zhang). In our opinion, it does not make sense to compare exact CIs with asymptotic CIs, since the former are obliged to respect the error $\alpha$ and the latter are not. On the other hand, we believe that the authors should have assessed their asymptotic Method (EC) in relation to two other asymptotic methods (MN and Wald) in a wide range of scenarios and not only in three examples. Indeed, the asymptotic CI for $P_T-P_C$ is a widely studied subject (Martín Andrés *et al.* 2012, Laud & Dane 2014, and Fagerland *et al.* 2015) and we would have expected the authors to have included some of the methods selected by other authors.

**Acknowledgments.-** This research was supported by the Spanish Ministry of Economy, Industry and Competitiveness under Grant PID2021-126095NB-I00 funded by MCIN/AEI/ 10.13039/501100011033 and by "ERDF A way of making Europe".